# LOW-MASS X-RAY BINARY POPULATIONS

G. Fabbiano[a]

[a]Harvard-Smithsonian Center for Astrophysics, 60 Garden St., Cambridge MA 02138, USA

Abstract. Low-mass X-ray binaries (LMXBs) have been studied in the Galaxy since the beginning of X-ray astronomy. A lot has been learned about these bright X-ray sources, but significant questions are still open. These questions are related to the origin and evolution of LMXBs - dynamical evolution in globular clusters (GC) or evolution of native field binaries -, and on how their properties may be related to their evolution. The discovery of several LMXB populations in elliptical galaxies with the *Chandra X-ray Observatory* provides new tools for studying these sources.

Keywords: X-ray binary stars

**PACS:** *97.80.Jp*

## 1.0 INTRODUCTION AND OUTLINE

The first population of extra-galactic LMXBs was discovered when the *Einstein Observatory* first imaged M31 in X-rays (van Speybroeck et al 1979). The presence of LMXB populations in elliptical galaxies was also inferred from the analysis of *Einstein* data (e.g., Trinchieri & Fabbiano 1985; Kim, Fabbiano & Trinchieri 1992b), but the richness of these populations and their importance for the overall X-ray emission of elliptical and S0 galaxies were controversial at the time (see Fabbiano 1989).

Sub-arcsecond resolution *Chandra* images have revealed rich populations of point-like sources in all elliptical galaxies at least as far as the Virgo cluster; some of these sources are found in GCs, others are in the stellar field of the parent galaxy (Sarazin, Irwin & Bregman 2000; see Fabbiano 2006). These sources have luminosities consistent with those of luminous X-ray binaries. They also have typically hard spectra, with possibly softer spectra in the most high luminosity sources, suggesting the presence of disk emission in black hole (BH) binaries (Irwin, Athey & Bregman 2003). The variability detected in cases of repeated *Chandra* observations is also consistent with the presence of compact accreting objects (Kraft et al 2001; Sivakoff, Sarazin & Jordan 2005; Brassington et al 2008, 2009).

In this paper I will review some of the recent *Chandra* results on LMXB population. Most, but not all, of the results presented here are based on the deep monitoring series of *Chandra* observations of the two nearby elliptical galaxies NGC 3379 and NGC 4278, which were obtained by a large collaboration of which I was PI. These two galaxies are virtually devoid of hot ISM, allowing the detection of point-like sources down to X-ray luminosities in the range observed in Galactic LMXBs. They have similar, old, stellar populations, but differ in the number of associated GCs, so to allow an observational study of the effect of GCs on the LMXB populations. The source catalogs derived from these observations can be found in Brassington et al (2008, 2009). Some of the material presented in this paper is also included in the proceedings of the lectures I gave at the 2009 Canary Islands Winter School (Fabbiano 2010).

The outline of this paper is as follows: in Section 2.0, I discuss the spectral variability patterns that link these sources to Galactic LMXBs; in Section 3.0, I give a brief overview of the properties of the X-ray luminosity function (XLF) of these LMXB populations; in Section 4.0, I revisit the question of LMXB origin and evolution in the light of these new results.

# 2.0 THE X-RAY SPECTRA OF THE LMXB POPULATIONS OF NGC 3379 AND NGC 4278

The deep monitoring observations of NGC 3379 and NGC 4278 give us the opportunity to study in details the spectra and the spectral variability of the most luminous X-ray sources in these galaxies, and to compare these results with the wealth of information that is available for Galactic LMXBs. Given their high luminosities, these sources are likely to be BH binaries (BHBs); does their spectral variability follow the patterns seen in Galactic BHBs (see review, Remillard & McClintock 2006)? In particular, can we distinguish between disk dominated (thermal, high/soft) states, power-law dominated (possibly jet-dominated) hard/low states, and steep power-law (very high; possibly Comptonized) states?

The X-ray spectra of Galactic BHBs are well represented with two spectral components, with varying flux ratios: a power-law and an accretion disk (see Remillard & McClintock 2006). The statistics of LMXB detection in external galaxies with *Chandra* do not give us enough counts to attempt meaningful double-component fits in most cases. Fitting the data with single-component models, either power-law or accretion disk (multi-temperature black body), typically yields comparable fit statistics. Therefore, we cannot discriminate between these models based on goodness of fit arguments alone. Is there some way of looking at the parameter spaces and use these results as a discriminator?

To explore how much is possible to learn from the spectral analysis of these data with single-component model fits, Brassington et al (2010) resorted to an extensive set of simulations, creating a range of two-component spectra and fitting the results to single-component models. The simulated spectra were power-law and accretion disk combinations, with the power-law photon index covering the range $\Gamma=1.5 - 2.3$ in steps of 0.2 and the disk temperature covering the range $kT = 0.5 - 2.25$ keV in steps of 0.25 keV. The flux ratios of the two components were varied between 0% and 100% for each combination of power-law photon index $\Gamma$ and disk temperature $kT$; in all cases only a Galactic line of sight neutral hydrogen absorption column $N_H$ was assumed. This $N_H$ is broadly speaking appropriate for LMXBs, where extensive stellar winds are not present; moreover, we wouldn't expect significant absorption within the elliptical galaxies, given the general lack of cold ISM in these galaxies. For each combination of parameters, 100 spectra were simulated with 500 and 1000 detected counts, with noise added.

This exercise showed that if the data are fitted with a single absorbed power-law model, the greater the contribution to the spectrum of a disk component, the larger (above the line of sight value) $N_H$ is returned by the fit. This effect is significant for the typical range of temperatures of Galactic LMXB accretion disks. Conversely, the presence of even a small power-law component in the spectrum yields un-physically small $N_H$, well below the line of sight value, when the spectrum is fitted with a single absorbed disk component (fig. 1).

These simulations demonstrate that the value of the best-fit $N_H$ can give indications on the relative prominence of the power-law or thermal emission in the spectra. With these results in mind, evaluating the spectra and spectral variability of the luminous LMXBs of NGC 3379 and NGC 4278 shows that the most luminous sources, and the variable sources when at the highest luminosity in these samples, favor more prominent accretion disk components (figs. 2 and 3; Brassington et al 2010; Fabbiano et al 2010). In disk-dominated spectra (selected according to the results of the simulations), flux-related temperature variations of the disk component are seen, which may be connected with variations in the accretion rate. Plotting these results in the BHB $L_X - T_{in}$ diagram, where $L_X$ is the source luminosity and $T_{in}$ is the temperature of the inner accretion disk, we find trends consistent with those observed in Galactic XRBs (fig. 4). Moreover, the majority of these sources are also consistent with BH masses in the range of those measured in Galactic BHBs (see Remillard & McClintock 2006; also Özel et al 2010, who show that the Galactic BH mass distribution is peaked at $7.8 \pm 1.2\ M_\odot$). These results confirm that luminous X-ray sources in elliptical galaxies are typical LMXB systems.

NGC 4278 yields sizeable samples of both field and GC LMXBs, allowing the comparison of the average spectral properties of selected subsamples (Fabbiano et al 2010). Using a single power-law model, the jointly fitted X-ray spectrum of the most luminous field LMXBs ($L_X > 1.5 \times 10^{38}$ erg s$^{-1}$) requires substantially larger $N_H$ than the jointly-fitted spectrum of less luminous field LMXBs, suggesting on average disk-dominated emission in these luminous sources. Comparison between average field and GC LMXB spectra show that high luminosity sources have compatible spectra, suggesting in both cases disk-dominated emission. However, lower-luminosity sources ($L_X < 6 \times 10^{37}$ erg s$^{-1}$) behave in an unexpected way. While field sources have an average power-law dominated spectrum with line of sight $N_H$, consistent with intrinsic power-law emission spectra, the GC LMXB average spectrum is either an absorbed power-law or a disk-dominated spectrum. This result opens the possibility that the populations of lower luminosity LMXBs in the field and in GCs may be intrinsically different.

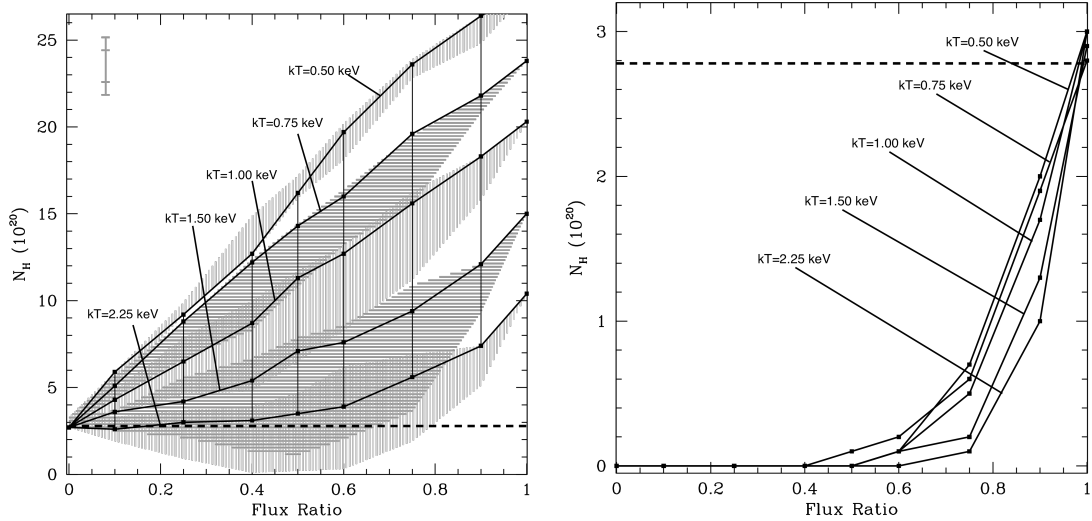

**FIGURE 1.** a) $N_H$ versus the ratio of disk/power-law flux for power-law only fits; solid lines are for $\Gamma=1.7$ power-laws; the horizontal dashed line is at the line of sight $N_H$. b) Same for disk only fits (Brassington et al 2010).

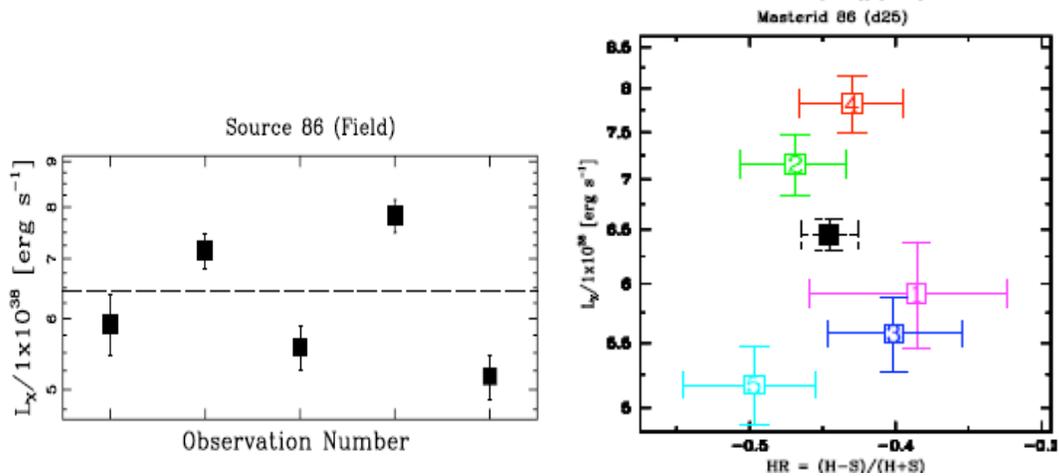

**FIGURE 2.** a) Luminosity of source 86 in NGC 3379 (Brassington et al 2008) in separate Chandra observations. b) Luminosity versus hardness ratio of this source, suggesting luminosity-correlated spectral variability.

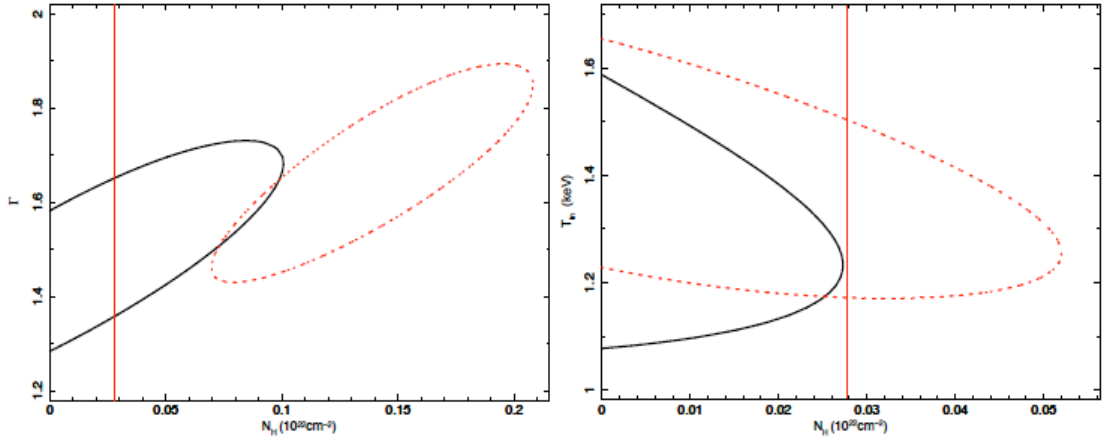

**FIGURE 3.** Results of the spectral fits of source 86 in NGC 3379. a) $\Gamma$-$N_H$ parameter space for single power-law fit; b) kT-$N_H$ parameter space for single disk fit. In both cases, the contours are $2\sigma$, solid observations 1+3, dashed observations 2+4, the vertical line represents the line of sight $N_H$. In the power-law fit the lower-luminosity spectra are consistent with Galactic line of sight $N_H$, while the high luminosity spectra exclude it; conversely the disk fit requires un-physically low $N_H$ in the low-luminosity spectrum, while the high luminosity parameter space is consistent with the line of sight $N_H$ (from Brassington et al 2010).

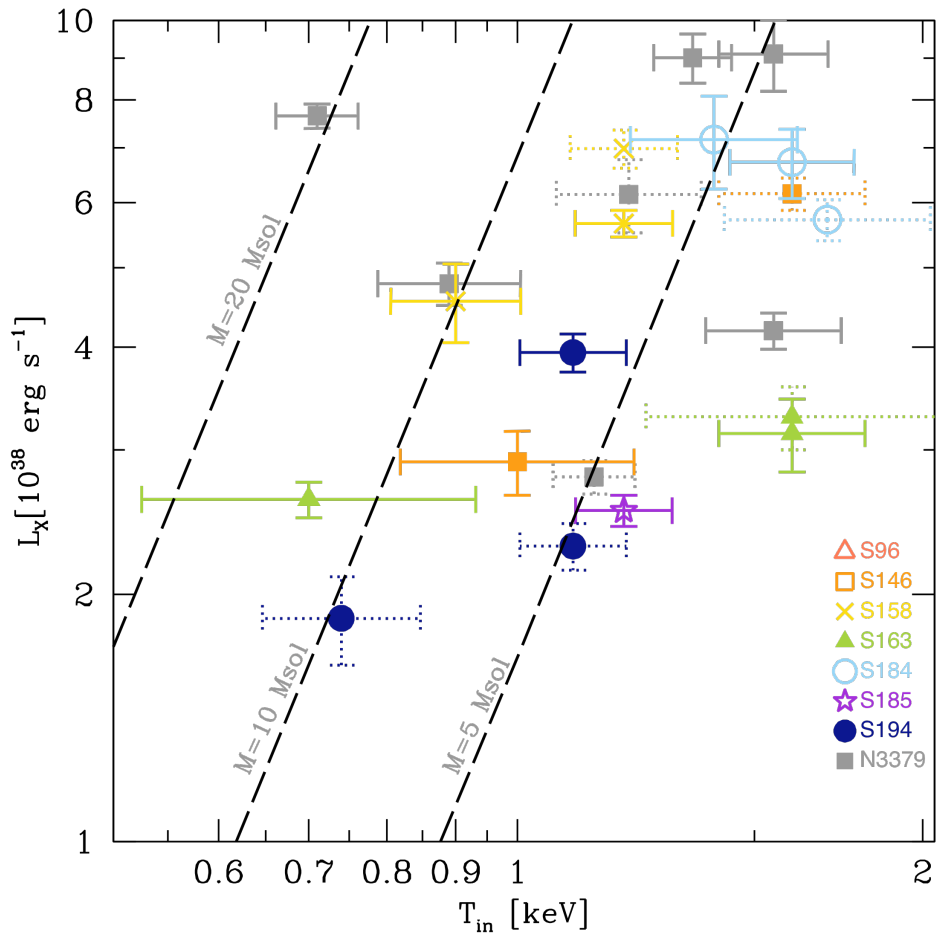

**FIGURE 4.** Luminosity –disk temperature diagram for sources in NGC 3379 and NGC4478 (Brassington et al 2010; Fabbiano et al 2010)

# 3.0 THE LMXB XLF

Several studies have addressed the LMXB XLF (see Fabbiano 2006 and refs. therein; also, see Voss, this book). The high luminosity ($L_X > 5 \times 10^{37}$ erg s$^{-1}$) LMXB XLF can be approximated with a steep power law with a high luminosity break (Kim & Fabbiano 2004; Gilfanov 2004). The deep images of NGC 3379, NGC 4278, augmented with NGC 4697 (PI: Sivakoff) have given us the means to explore the lower luminosity XLF. The questions we sought to address were: (1) How does the XLF behave at the lower X-ray luminosities more typical of Galactic LMXBs? In particular, is there a low luminosity break, as suggested by the Galactic LMXB XLF (Grimm, Gilfanov & Sunyaev 2002)? (2) In the high luminosity regime no differences were found between field and GC LMXB XLFs (Kim E. et al 2006). This similarity is consistent with (but does not prove) a similar evolutionary path for the two LMXB populations. Does this similarity extend to lower LMXB luminosities?

Fig. 5 gives a summary of the present understanding, based on the work by Kim et al (2009) and Kim & Fabbiano (2010).

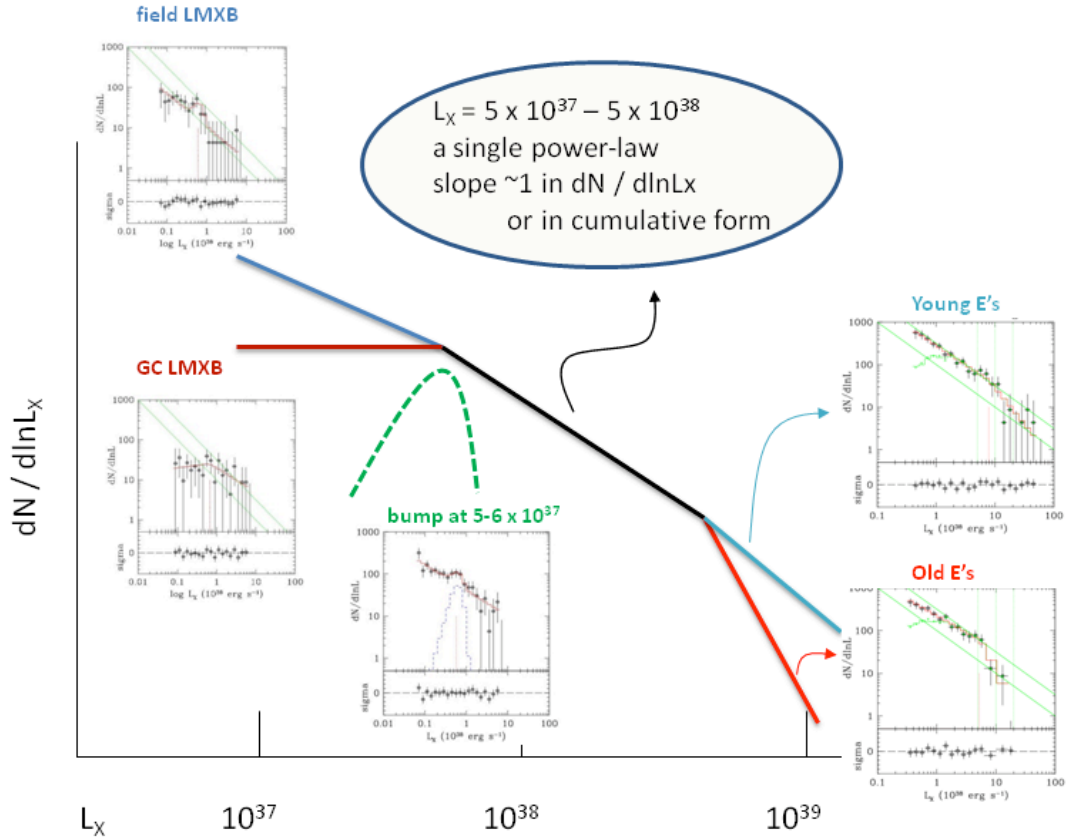

**FIGURE 5.** Schematic summary of the LMXB XLF features (Kim & Fabbiano 2010)

The deep *Chandra* observations of the three elliptical galaxies NGC 3379, NGC 4278 and NGC 4697 confirm the presence of a low-luminosity break, leading to a flatter XLF at luminosities $<5\times10^{37}$ erg s$^{-1}$ (Kim et al 2009). This lack of low luminosity LMXBs is more pronounced in the LMXBs associated with GCs.

There are at least three papers in the literature attempting to explain this break. Postnov & Kuranov (2005) suggest that the break may be related to the change of the binary braking mechanism from magnetic stellar winds at high luminosity, to gravitational radiation at low luminosity. Revnivtsev et al (2008) instead propose that the break may reflect the transition from the high/soft disk dominated emission, where the luminosity is proportional to the accretion rate, to the low/hard state regime, where for advection dominated flows the luminosity follows a steeper function of the accretion rate; assuming an intrinsic relation between the number of detected LMXBs and the accretion rate, they can reproduce the XLF break. A different approach was followed by Kim et al (2009), who compared the XLF of field-LMXBs with the PS model of Fragos et al (2008), suggesting that the break may be a 'bump', related to contribution of the Red Giant LMXB population.

Kim et al (2009) also sought to explain the dearth of GC LMXB at $L_X < 5\times10^{37}$ erg s$^{-1}$, relative to the field LMXB population (see fig. 5). It is unlikely that this result is due to underestimating the number of low-luminosity GC LMXBs because of source confusion, resulting in 'spurious' detections of high-$L_X$ LMXBs. Even if crowding may occur (see Section 4.1), the variability pattern of GC sources (Brassington et al 2008, 2009) suggests that it is unlikely that this is an overwhelming effect. Another possibility is that transient sources may appear prevalently in the field population, thereby spuriously steepening the field LMXB XLF at the high luminosity (so that the perceived lack of low-luminosity GC sources would instead be an excess of high-luminosity GC sources). However, some instances of transients (or possible transients) have also been reported in GCs (Brassington et al 2008, 2009). The possibility that the break may be related to the onset of transient behavior in ultra-compact binaries (Neutron star +White dwarf) that could be responsible for the GC X-ray emission (Bildsten & Deloye 2004) has been debated, but not conclusively because the break in that case should appear at even lower luminosity. It is clear that this is an area where more theoretical work will be needed to explain satisfactorily the observations.

Interesting variations in the slope of the XLF are also found at the high luminosity end ($L_X > 5\times10^{38}$ erg s$^{-1}$). These differences are correlated with the age of the parent stellar population. Although the stellar population of elliptical galaxies is old, there are galaxies where either morphological or spectral indicators suggest relatively recent merging interaction and rejuvenation (see Schweizer & Seitzer 1992). Early reports of rejuvenation in the LMXB population noted asymmetrically distributed and luminous sources in NGC 720 (Jeltema et al 2003) and NGC 4261 (Zezas et al 2003). In a recent work Kim & Fabbiano (2010) have looked at two samples of 'rejuvenated' and 'old' elliptical galaxies observed with *Chandra*, finding a difference in the luminosity functions that demonstrates an excess of very luminous sources ($L_X > 5\times10^{38}$ erg s$^{-1}$) in the 'rejuvenated' galaxies. This feature provides an observational diagnostic of rejuvenation in early-type galaxies.

## 4.0 LMXB ORIGIN AND EVOLUTION

So far, I have discussed the different shapes of the LMXB XLF. But information is also given by the normalization, which is a measure of the overall number of LMXBs in a given galaxy. Gilfanov (2004) pointed out that this normalization should be a function of the total stellar mass of the bulge, given that LMXBs are long-lived systems. This conclusion is consistent with the earlier suggestion of Fabbiano & Trinchieri (1985), which was based on the presence of a correlation between the X-ray and H-band near-IR luminosities in bulge-dominated galaxies observed with *Einstein*. However, it was also pointed out that the specific frequency of GCs (number of GC per unit galaxy stellar mass) in a galaxy would affect the number of LMXBs, especially if dynamical formation is the dominant mechanism (White et al 2002). Kim & Fabbiano (2004) showed that the XLF normalization depends on both total stellar luminosity/mass and GC specific frequency, suggesting that dynamical formation in GC with subsequent dispersion in the stellar field (see Clark 1975; Grindlay 1984) and the evolution of native binaries (see Verbunt & van den Heuvel 1995) may both have a place in the formation of LMXBs.

The deep *Chandra* observations of the three elliptical galaxies NGC 3379, NGC 4278 and NGC 4697, providing source counts and optical identification down to similarly deep detection thresholds, were used to revisit the question of field evolution versus GC formation of LMXBs. Kim et al (2009) find that the number of GC LMXBs increases with the galaxy GC specific frequency ($S_N$), as expected. However, the number of field LMXBs also increases with $S_N$, albeit with a weaker dependence (fig. 6). This result suggests that while there is a fraction of LMXBs, which derive from the evolution of native field binaries (see also Irwin 2005, Juett 2005, Kim E. et al 2006), the field-LMXB population contains also sources that were formed in GCs.

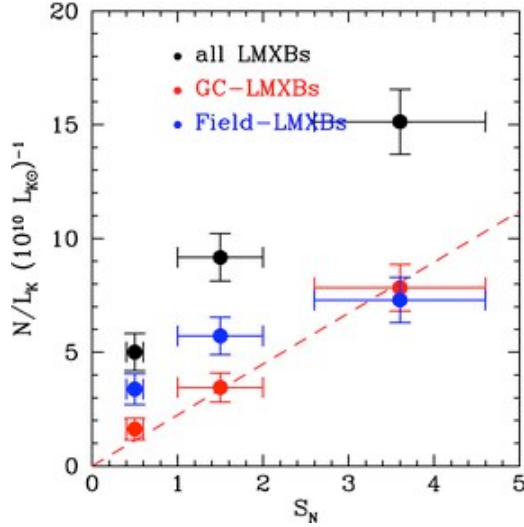

**FIGURE 6**. Number counts (per unit galaxy K-band luminosity) of LMXBs detected in three galaxies versus the GC specific frequency. The number of GC LMXBs increases with the GC specific frequency, as expected, but there is also a similar – although smaller – effect in the field-LMXBs (Kim et al 2009)

## 4.1 LMXBs and GCs: Properties and Correlations

Several studies have attempted to explore the LMXB-GC connection by combining the samples of LMXBs from *Chandra* observations with samples of GCs detected with *HST*. These works have led to the identification of field and GC LMXB samples, which have then been compared in the attempt to set observational constraints to the evolution and nature of these LMXBs (see e.g., Fabbiano 2006; Brassington et al 2008, 2009; Maccarone, Kundu & Zepf 2007; Sivakoff et al 2007).

Three key factors have emerged, which favor the presence of a LMXB in a GC: GC luminosity, metallicity, and compactness (e.g., Kundu, Maccarone & Zepf 2002; Maccarone, Kundu & Zepf 2003; Kim E. et al 2006; Sivakoff et al 2007). The connections with luminosity and compactness link directly the presence of LMXBs with an environment that fosters their dynamical formation. The reason for the preference of red, high-metallicity GCs as LMXB hosts is less clear. Stellar winds speeding up the LMXB evolution in blue metal-poor GCs have been suggested as a way to eliminate LMXBs in these clusters (Maccarone, Kundu & Zepf 2004). Another mechanism, proposed by Ivanova (2006), is the lack of outer convective zones in metal-poor main sequence stars resulting in lack of magnetic braking, and thus impeding the formation of a compact binary system in blue GCs.

In the Milky Way, where one can probe the X-ray emission of GCs down to a very low luminosity ($\sim 10^{30}$ erg s$^{-1}$ with *Chandra*, e.g., Grindlay et al 2001), denser and more compact GCs tend preferentially to host LMXBs; these are the GCs where dynamical interactions would be favored (Verbunt & Lewin 2006; Bregman et al 2006). These more compact GCs are found at smaller galactocentric radii ($R_G$), independent of their luminosity (van den Bergh, Morbey & Pazder 1991). In their *Chandra* and *HST* survey of Virgo galaxies Sivakoff et al (2007) did not find any $R_G$ dependence of the association of an X-ray source ($L_X > 5 \times 10^{37}$ erg s$^{-1}$), with a GC. They instead report at all radii strong dependences on GC luminosity, compactness and color. Using our deeper *Chandra* survey of NGC 4278, we find a significant radial effect - not on the presence - but on the luminosity of the X-ray sources (Fabbiano et al 2010): the more luminous X-ray sources tend to be found at smaller $R_G$ in this galaxy (fig. 11). A similar effect is not found in the control sample of field LMXBs in NGC 4278, ruling out analysis and sampling biases and pointing to an intrinsic property of GC LMXBs.

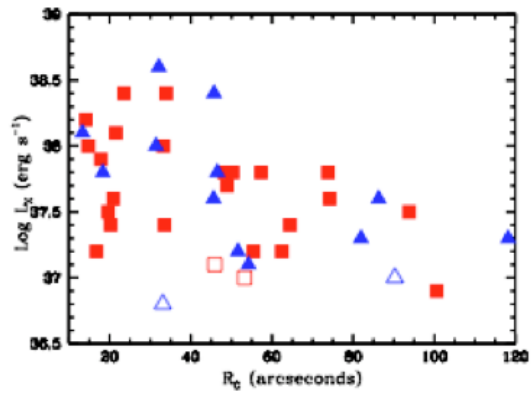

**FIGURE 7.** X-ray luminosity of GC sources in NGC 4278 versus RG. Red-GC are squares and blue-GC triangles. The Spearman Rank probability of no correlation is P=0.0008 (Fabbiano et al 2010)

As observed in the Milky Way (van den Bergh et al 1991), there is no $R_G$ dependence of the GC luminosity in NGC 4278, while there is a very weak link between GC luminosity and X-ray source luminosity. If GCs at smaller $R_G$ are more compact, as in the Milky Way, the NGC 4278 result may be explained with a higher probability of LMXB formation in these GCs, resulting in multiple luminous LMXBs in a single cluster. The higher X-ray luminosities of the more central GCs could be due to the presence of multiple LMXBs. In the Milky Way the number of GC X-ray sources increases with a higher formation rate parameter (a function of both GC luminosity and compactness; see Pooley et al 2003); lacking a radial dependence of GC luminosity, GC compactness would be mostly responsible for the larger X-ray luminosity of more central GCs in NGC 4278.

## ACKNOWLEDGMENTS

I thank Nicky Brassington and Dong-Woo Kim for providing some of the material and figures in this paper. Nicky Brassington produced the extensive simulations needed for interpreting the results of the spectral fits discussed here. I acknowledge the *Chandra X-ray Center* for partial support under NASA contract NAS8-39073.